\def\binom#1#2{\left( \begin{array}{c} #1 \\ #2 \end{array}\right)}

\documentstyle[preprint,aps,epsf,floats,tighten]{revtex}

\begin{document}

\preprint{t97/074 ; cond-mat/9707023}

\title{Asymptotic behavior of  \\ two-terminal series-parallel
networks}

\author{ O. Golinelli\thanks{email: golinelli@cea.fr}}

\address{ CEA Saclay, Service de Physique Th\'eorique,
\\ F-91191 Gif-sur-Yvette, France}

\date{\today}
\maketitle

\begin{abstract}

This paper discusses the enumeration of two-terminal series-parallel
networks, i.e. the number of electrical networks built with $n$
identical elements connected in series or parallel with two-terminal
nodes.  They frequently occur in applied probability theory as a model
for real networks.  The number of networks grows asymptotically like
$R^n/n^{\alpha}$, as for some models of statistical physics like
self-avoiding walks, lattice animals, meanders, etc.  By using a exact
recurrence relation, the entropy is numerically estimated at $R =
3.560\ 839\ 309\ 538\ 943\ 3\ (1)$, and we show that the sub-leading
``universal'' exponent $\alpha$ is 3/2.

\end{abstract}
\pacs {}

\narrowtext
\section{Introduction}

The series-parallel networks are one of the simplest models of real
networks.  Then even if it is at last one century old (Mac Mahon 1892),
it still occurs frequently in both theoretical or applied studies.
Some questions like the reliability (Feo and Johnson 1990), the
survival probability after failures, the routing problem and the
queuing delay (Tamir 1993), etc.\ are now an important branch of
probability theory applied to communication, computer or power
networks.

As usual in this kind of model, the elementary components (resistances
or switches) are chosen all identical.  Then the interest is in the
behavior of the structure made of series or parallel connections.

Some authors (Riordan 1958, Lomnicki 1972) investigated this problem
with a combinatorial point of view, and, in particular, give
recurrences to compute the number of different structures built with
$n$ elements.  Then, they consider the problem as formally ``solved''.

In this article, we study this model by using analogies with
statistical physics models like self-avoiding walks (Des Cloizeaux and
Jannink 1987),  lattice animals and percolation (Derrida and De Seze
1982), folding and meanders model (Di Francesco et al. 1996), etc.  In
all these models, the number of configurations for $n$ elements (steps,
links, folds) grows asymptotically like $R^n / n^\alpha$ where $R$
depends on the details of the model and $\alpha$ is an ``universal''
exponent, common to a broad set of models (``universality'' class).

In Section 2, we define the problem.  In Section 3, we explain how to
enumerate the series-parallel networks.  Finally, in Section 4, we give
by numerical extrapolation a value for $R$ and we show that $\alpha =
3/2$.

\section{Definitions and notations}

The two-terminal series-parallel networks are electric network
considered geometrically.  They are built with {\sl elements} $X_i$
which are conductor ($X_i = 1$) or insulator ($X_i = 0$).  When two
structures (with conductivities $X_a$ and $X_b$) are in parallel, their
(global) conductivity is $X_a + X_b$ where $+$ is the boolean operator
{\sl Or}.  When they are in series, it is $X_1 X_2$ where the product
is the boolean operator {\sl And}.

A network with $n$ elements, which connects two-terminal nodes, has a
boolean representation $F = F(X_1,\dots,X_n)$.  We consider that two
networks are {\sl equivalent} if their boolean representations differ
only by a permutation of variables $\{X_i\}$.  For example, $X_1 X_2
(X_3 + X_4)$, $X_1 (X_2 + X_3) X_4$ and $(X_1 + X_2) X_3 X_4$ are
equivalent.  We note that this definition is less restrictive than
topological definition.  In other words, equivalence consists of
interchange inside a group of structures in series, or in parallel.
The networks for $n=1,2,3,4$ are given in Table~\ref{ta}.

\begin{table}
\caption {Two-terminal series-parallel networks for $n=1,2,3,4$.  On
each row, the two networks are dual.  The $a_n$'s (number of networks)
are 1, 2, 4, 10, \dots  The $b_n$'s (number of essentially series or,
by duality, essentially parallel networks) are 1, 1, 2, 5, \dots}
\label{ta}
\begin{tabular}{cclclc}
number of & essentially & & essentially & & number of \\
elements  & series      & & parallel    & & networks \\
\hline
1 & \epsfbox{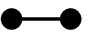} & $X_1$ & \epsfbox{11.eps} & $X_1$ & 1\\
\hline
2 & \epsfbox{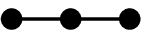} & $X_1 X_2$ & \epsfbox{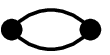} & $X_1 + X_2$ & 2 \\
\hline
3 & \epsfbox{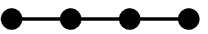} & $X_1 X_2 X_3$ &
    \epsfbox{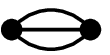} & $X_1 + X_2 + X_3$ & 4\\
  & \epsfbox{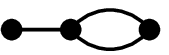} & $X_1 (X_2 + X_3)$ &
    \epsfbox{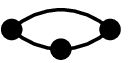} & $X_1 + X_2 X_3$ & \\
\hline
4 & \epsfbox{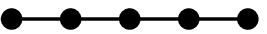} & $X_1 X_2 X_3 X_4$ &
    \epsfbox{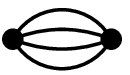} & $X_1 + X_2 + X_3 + X_4$ & 10\\
  & \epsfbox{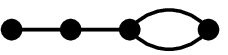} & $X_1 X_2 (X_3 + X_4)$ &
    \epsfbox{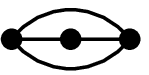} & $X_1 + X_2 + X_3 X_4$ & \\
  & \epsfbox{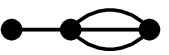} & $X_1 (X_2 + X_3 + X_4)$ &
    \epsfbox{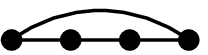} & $X_1 + X_2 X_3 X_4$ & \\
  & \epsfbox{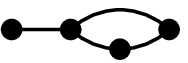} & $X_1 (X_2 + X_3 X_4)$ &
    \epsfbox{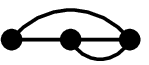} & $X_1 + X_2 (X_3 + X_4)$ & \\
  & \epsfbox{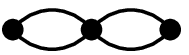} & $(X_1 + X_2) (X_3 + X_4)$ &
    \epsfbox{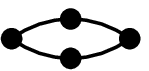} & $X_1 X_2 + X_3 X_4$ & \\
\end{tabular}
\end{table}

Let us call $a_n$ the number of different networks with $n$ elements.
It is also the number of non-equivalent boolean functions of $n$
variables, built only with {\sl And} and {\sl Or}.

Except for the trivial case $n=1$ with $a_1=1$, we see in
Table~\ref{ta} two classes of networks~: essentially series
($\sigma$-{\sl networks}) and essentially parallel ($\pi$-{\sl
networks}).  There is a {\sl duality} :  by changing the boolean $\sl
And$ into {\sl Or} and vice versa, each $\sigma$-network is dual of a
$\pi$-network.  Let us call $b_n$ the number of $\sigma$-networks (or
of $\pi$-networks).  We have
\begin{equation}
a_n = 2 b_n - \delta_{n,1}.
\end{equation}

\section{Enumeration and recurrence generating function}

In this section, we explain how to enumerate the different networks,
i.e.  compute the $a_n$ or $b_n$.  By use of the duality, we consider
only a $\sigma$-network $S_n$ with $n \ge 2$ elements.  It is composed
with $k$ $\pi$-structures of $w_1,w_2\dots,w_k$ elements, with $k \ge
2$ and $\sum_i w_i = n$.  Its boolean representation is $ S_n =
P_{w_1}P_{w_2}\dots P_{w_k}$.  By grouping together the equal $w_i$, we
can use a symbolic exponential notation $S_n = P_1^{r_1} P_2^{r_2}
\dots P_q^{r_q}$.

The $\{r_j\}$ are non-negative ($r_j \ge 0$) integers~; they form a
{\sl partition} of $n$ with at least two parts~: $n = \sum_j j\;r_j$ is
partitioned in $k = \sum_j r_j$ parts.  We note a partition as $1^{r_1}
2^{r_2} \dots q^{r_q}$.  For example, the partitions of 5 are $\{1^5,
1^32, 1^23, 12^2, 14, 23\}$.

Then, to obtain $b_n$ (the number of $\sigma$-network with $n$
elements), we first must compute the contribution of a given partition,
and finally, we will sum over all the partitions.  Thanks to the above
decomposition, the number of $S_n$ with a given partition $\{r_j\}$ is
a product of as many factors as $j$, dependent only on $b_j$ and
$r_j$.

For example, it is obvious that, for $b_5$, the contribution of the
partition $(23)$ is $b_2b_3$, because $b_2$ is the number of choices
for the first $\pi$-structure of two elements, and $b_3$ for the second
of three elements.

When two $\pi$-structures have the same number of elements, i.e.
$r_j=2$, their contribution is not $b_j^2$ but, to remove duplications
obtained by permutations, $b_j (b_j+1)/2$.  More generally, for
arbitrary $r_j \ge 0$, the factor is
\begin{equation}
\binom{b_j+r_j-1}{r_j}
= \sum_{1\le u_1 \le u_2 \dots \le u_{r_j} \le b_j} \!\!\!\!\!\! 1 .
\end{equation}
So, the $b_n$ (or the $a_n$) can be computed by recurrence
\begin{equation}
\label{b_n}
b_n = \sum_{\{r_j\}} \prod_j \binom{b_j+r_j-1}{r_j}
\end{equation}
where the sum runs over all the partition $\{r_j\}$ of $n$ with at
least two parts, as explained above.  For example, $b_5 = 1 + b_2 + b_3
+ b_2(b_2+1)/2 + b_4 + b_2b_3 = 12$.

With a computer, the $b_n$ can be calculated by recurrence.
Unfortunately, the number of partitions of $n$ grows like $\exp(\pi
\sqrt{2n/3}) / n$ (Hardy-Ramanujan formula).  We will now describe how
to obtain a more practical recurrence formula.

We will use the generating functions
\begin{eqnarray}
b(t) &=& \sum_{i\ge 1} b_i t^i = t + t^2 + 2t^3 + 5 t^4 + \dots, \\
a(t) &=& \sum_{i\ge 1} a_i t^i = t + 2t^2 + 4t^3 + 10 t^4 + \dots
= 2 b(t) - t.
\end{eqnarray}
Following an argument similar from that used by Euler for the
generating function for the partition of $n$, by using Eq.~(\ref{b_n})
\begin{eqnarray}
&&(1 + \binom{b_1}{1} t + \binom{b_1+1}{2} t^2 + \binom{b_1+2}{3} t^3 +
\dots) \nonumber \\
&\times & (1 + \binom{b_2}{1} t^2 + \binom{b_2+1}{2} t^4 +
\binom{b_2+2}{3} t^6 + \dots) \nonumber \\
&\times & \dots \nonumber \\
&\times & (1 + \binom{b_j}{1} t^j + \binom{b_j+1}{2} t^{2j} + \dots +
\binom{b_j+r-1}{r} t^{rj} + \dots) \label{gene}\\
&\times & \dots \nonumber \\
&= & 1 + \sum_{n \ge 1} \binom{b_n}{1} t^n + \sum_{n \ge 2} b_n t^n
= 1 + a(t).
\end{eqnarray}
As a generic line of the l.h.s.\ of Eq.~(\ref{gene}) is equal to
$(1-t^j)^{-b_j}$, we obtain a formula attributed to Mac Mahon (1892)~:
\begin{equation}
\label{mahon}
1+a(t) = \prod_{j\ge 1} (1-t^j)^{-b_j}
\end{equation}
Eq~(\ref{mahon}) can be rewritten as a non-local functional relation
\begin{equation}
\ln[1+a(t)] = \sum_{k\ge 1} \sum_{j\ge 1} {1 \over k} b_j t^{kj}
= \sum_{k\ge 1} {1\over k} b(t^k).  \label{jat} 
\end{equation}
We obtain a useful recurrence taking the derivative of Eq.~(\ref{jat})
\begin{equation}
t {a'(t) \over 1+a(t)} = \sum_{k\ge 1} \sum_{j\ge 1} j b_j t^{kj}
\end{equation}
Defining $e_i = \sum_{j|i} j a_j$, where the sum runs over the divisors
of $i$ including $j=1$ and $j=i$,
\begin{equation}
2 \sum_{n\ge 1} n a_n t^n = (1+\sum_{j\ge 1} a_j t^j) \sum_{i\ge 1}
t^i (1+e_i)
\end{equation}
Defining $e'_i = \sum_{j|i ; j \neq i} j a_j = e_i - i a_i$, we can
compute the $a_n$ with the recurrence formula (Lomnicki 1972)
\begin{equation}
\label{lomni}
n a_n = 1 + e'_n + \sum_{i=1}^{n-1} a_{n-i} (1 + e_i).
\end{equation}
for which the computational effort grows like $n^2$.  By programming
this recurrence on a computer with the use of {\sl real*16} (i.e. real
numbers described by 128 bits), it is possible to obtain several
thousands of terms with 30 decimal digits of precision.  It is more
than enough to determine the asymptotic behavior.

This integer sequence belongs to the encyclopedia of Sloane (1996).
The beginning is given in Table~\ref{li}.

\begin{table}
\caption {The $a_n$ for $n \le 45$}
\label{li}
\begin{tabular}{rrrrrr}
$n$ & $a_n$ & $n$ & $a_n$ & $n$ & $a_n$ \\
\hline
1 & 1 & 16 & 4 507 352 & 31 & 306 481 637 180 676\\
2 & 2 & 17 & 14 611 576 & 32 & 1 039 825 579 149 516\\
3 & 4 & 18 & 47 633 486 & 33 & 3 533 238 105 431 144\\
4 & 10 & 19 & 156 047 204 & 34 & 12 022 695 376 246 364\\
5 & 24 & 20 & 513 477 502 & 35 & 40 964 814 174 562 504\\
6 & 66 & 21 & 1 696 305 728 & 36 & 139 754 932 965 689 486\\
7 & 180 & 22 & 5 623 993 944 & 37 & 477 353 256 699 920 544\\
8 & 522 & 23 & 18 706 733 128 & 38 & 1 632 304 839 666 178 522\\
9 & 1 532 & 24 & 62 408 176 762 & 39 & 5 587 602 984 719 422 092\\
10 & 4 624 & 25 & 208 769 240 140 & 40 & 19 146 480 336 548 243 722\\
11 & 14 136 & 26 & 700 129 713 630 & 41 & 65 670 379 223 051 730 824\\
12 & 43 930 & 27 & 2 353 386 723 912 &
                                      42 & 225 448 266 880 936 115 440\\
13 & 137 908 & 28 & 7 927 504 004 640 &
				      43 & 774 644 024 342 635 183 112\\
14 & 437 502 & 29 & 26 757 247 573 360 &
				    44 & 2 663 899 039 964 678 479 766\\
15 & 1 399 068 & 30 & 90 479 177 302 242 &
				      45 & 9 168 056 380 423 173 365 752
\end{tabular}
\end{table}

\section{Asymptotic behavior}

A brief investigation of Table~\ref{li} shows that $a_{n+1} \approx 3.5
\ a_n$. Then, we can define a non-vanishing entropy per element $\ln
a_n / n \sim \ln R$. More generally, it is expected that the
exponential dominant behavior is corrected by sub-leading power-law for
large $n$
\begin{equation}
\label{ra}
a_n \sim C {R^n \over n^\alpha}
\end{equation}
where $C$ is a constant.

In fact, in Eq.~(\ref{lomni}), for large $n$, $e_n \sim n a_n$, then
$a_n \sim \sum_{i=1}^{n-1} a_i a_{n-i}$.  This last relation is
identical apart from a shift $(n \rightarrow n+2)$ to the definition
$c_{n+1} \equiv \sum_{j=0}^n c_j c_{n-j}$ of the Catalan numbers $c_n =
(2n)! / [n! (n+1)!]$.  Using the Stirling formula, we see that $c_n$
have the asymptotic behavior of Eq.~(\ref{ra}) with $R = 4$ and $\alpha
= 3/2$.  The fact that the $a_n$ and $c_n$ have the same asymptotic
relation do not imply that they have the same $R$.  As $\alpha > 0$, in
the sum, the extremal terms are dominant~: $c_0 c_n > c_1 c_{n-1} > c_2
c_{n-2} > \dots$.  Then the value of $R$ strongly depends of the value
of the first $c_n$ ($n$ small) and the exponential increase of $a_n$
and $c_n$ are not described by the same $R$.  On the other hand, we
expected the same $\alpha$ as we see below.

This behavior is common in statistical physics.  The enumeration of
random walks, self-avoiding walk, animals, etc.\ of $n$ step (or links)
drawn on a regular lattice always exhibits a exponential increase $R^n$
where $R$ depend on the connectivity of the lattice, and a sub-leading
$n^{-\alpha}$, where $\alpha$ is ``universal'', i.e. is dependent on
the physical nature of the problem (dimension, phases, \dots) but
independent on the details of the lattice.

Let us take a big two-terminal series-parallel network.  At each scale,
it is ``decorated'' by small parts $b_n$ ($n = 1,2, \dots$).  Then, the
contribution of these small $b_n$ is a large fraction of the total
entropy $\ln R$ ; they are similar to the connectivity of the lattice
and govern the dominant behavior $R^n$.  Nevertheless, the sub-leading
$n^{-\alpha}$ depends on more general properties and $\alpha$ can be
see as an ``universal'' exponent.  It described the singularity of the
generating function.

We suppose that the $a_n$ have the following large $n$ expansion
\begin{equation}
a_n \sim C {R^n \over n^{\alpha}} \ ( 1 + {u_1 \over n}
+ {u_2 \over n^2} + \dots )
\label{expa}
\end{equation}
and we want estimate the values of $R$ and $\alpha$ by numerical
extrapolation.  It is more convenient to study $V_n$ and $W_n$
\begin{eqnarray}
V_n & \equiv & \ln {a_n \over a_{n-1}}  =  \ln R - {\alpha \over n} +
{v_2 \over n^2} + {v_3 \over n^3} + \dots \\
W_n & \equiv & n^2 \ln {a_n a_{n-2} \over a^2_{n-1}}  = 
\alpha + {w_1 \over n} + {w_2 \over n^2} + \dots
\end{eqnarray}
By plotting $V_n$ (resp. $W_n$) {\sl versus} $1/n$, the
hypothesis~(\ref{expa}) is consistent if the data draw a straight
line~; its intersection with the y-axis gives $\ln R$ (resp.
$\alpha$).

A more systematic scheme of extrapolation is the Richardson polynomial
extrapolation process (Brezinski and Redivo Zaglia 1991).  Supposing
Eq.~(\ref{expa}),
\begin{eqnarray}
{1 \over p!} \Delta^p (n^p V_n) & = & \ln R + O(1/n^{p+1}) \label{dvn}\\
{1 \over p!} \Delta^p (n^p W_n) & = & \alpha + O(1/n^{p+1}) \nonumber
\end{eqnarray}
where $\Delta$ is the difference operator $\Delta f_n \equiv f_n -
f_{n-1}$.  This method gives the limits ($R$ and $\alpha$) and show
whether the hypothesis~(\ref{expa}) is consistent or not.  When $p$ is
large, the acceleration of convergence is faster~; unfortunately, the
numerical instabilities are amplified and after a given $n$,
extrapolated values (Eq.~\ref{dvn}) become chaotic.  In this work,
better extrapolations, obtained with $p$ between 3 and 5, are
\begin{eqnarray}
R &=& 3.560\ 839\ 309\ 538\ 943\ 3 \pm 10^{-16}, \\
\alpha &=& 1.5 \pm 10^{-13}.
\end{eqnarray}

\begin{figure}
\centering\leavevmode
\epsfxsize=10cm
\epsfbox{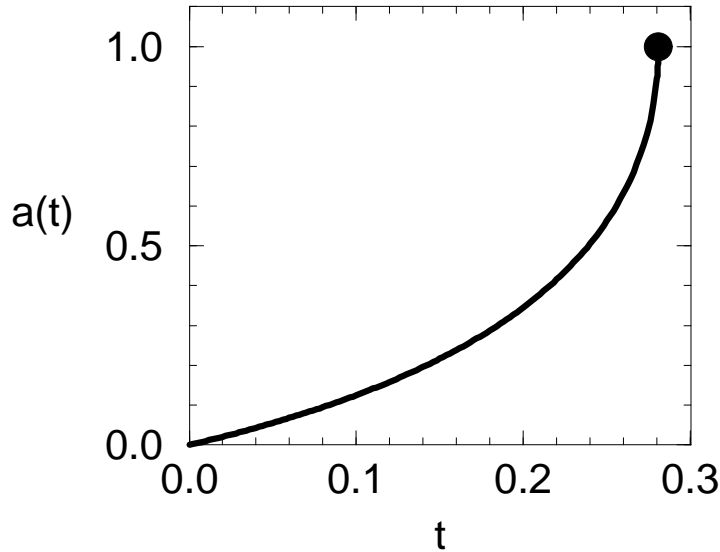}
\caption{The generating function $a(t)$.  Its singularity occurs at
$t^*=1/R \approx 0.28$.  At this point, $a(t^*)=1$ and the function has
a square root behavior because $\alpha = 3/2$.}
\end{figure}

We will now prove that $\alpha = 3/2$, by examining the behavior of the
generating function, which is numerically plotted on Fig.~1.  Clearly,
the radius of convergence of the series $a(t) = \sum_{i\ge 1} a_i t^i$
(i.e. the location of its singularity) is $t^* = 1/R$.  At $t=t^*$,
$a(t^*) \sim \sum_i 1/i^\alpha$ converges if $\alpha > 1$, and its
first derivative $a'(t^*) \sim \sum_i 1/i^{\alpha -1}$ diverges if
$\alpha < 2$.  Then, Fig.~1 indicates that $1<\alpha <2$.

More generally, the behavior of $a(t)$ around $t^*$ is given by
$\alpha$.  Let us consider the singular part $y(\epsilon) = a(t^*) -
a(t)$ with $t = t^* e^{-\epsilon} $. For small $\epsilon$, $y(\epsilon)
\sim \sum_i [1-e^{-\epsilon i}] / i^\alpha$.  For $\alpha > 2$, the sum
is dominated by small $i$ and $y(\epsilon) = O(\epsilon)$, i.e. the
first derivative $a'(t^*)$ is finite.  For $1<\alpha<2$, the sum is
dominated by $i = O(1/\epsilon)$ and $y(\epsilon) =
O(\epsilon^{\alpha-1})$.  The case $\alpha = 3/2$ corresponds to
$y(\epsilon) = O(\sqrt\epsilon)$.

Rewriting Eq.~(\ref{jat})
\begin{equation}
\ln [1+a(t)] = {1\over 2} [a(t) + t] + r(t) \ \ \  {\rm with}  \ \ \ 
r(t) = \sum_{k \ge 2} {1 \over 2k} [a(t^k) + t^k],
\end{equation}
for $t \le t^*$, $r(t)$ is a regular and increasing function.  By
defining $g(x) = 2\ln(1+x) -x$, $t+2r(t) = g(a(t))$.  The function
$g(x)$ has a maximum at $x^*=1$ with $g(x^*) = 2\ln 2-1$ and a
parabolic shape (i.e. a finite and negative second derivative).  As
$a(t) = g^{-1}(t+2r(t))$, this gives a singularity for
\begin{equation}
t^* + 2r(t^*) = 2\ln 2 -1 \ \ \ {\rm and} \ \ \ a(t^*) = 1
\end{equation}
with a square root shape $a(t^*)-a(t) = O(\sqrt{t^*-t})$.  This proves
that $\alpha = 3/2$.  Unfortunately, with this argument, we can not
compute the value of $t^* = 1/R$, because of the complexity of $r(t)$.
Then $R$ is sensitive to the details of the regular part of the
recurrence~: $R$ is not ``universal''.  But $\alpha$ depends only of
the kind of singularity : it is an ``universal'' exponent.  As we have
seen above, it is not surprising that it is equal to the $\alpha$ of
the Catalan numbers.

\section{Conclusion}

We have seen that some concepts of statistical physics can be
transposed to a problem of graph theory.  The partition function of
two-terminal series-parallel networks have an asymptotic behavior
similar to those of self-avoiding walks or lattice animals.  Its
exponential growth depends on details on the model.  On the other hand,
a power-law correction, with an ``universal'' exponent can be computed
exactly and have a fractional value $-3/2$.  For the self-avoiding
walks, this exponent is sensitive to the dimension and the kind of
boundaries conditions and others independent exponents are related to
geometrical properties.  In our problem, it will be interesting to have
a similar interpretation.  For example, the number of terminal nodes
(two in this article) could play the role of boundary conditions.  By
studying quantities like the average depth of elements (i.e. the number
of decomposition needed to isolate this element), it is probable that
new exponents would appear.

\acknowledgements

We thank P. Brax for a careful reading of the manuscript.


\end{document}